\documentclass[%
 aip,
rsi,%
 amsmath,amssymb,
 reprint,%
]{revtex4-1}

\usepackage{graphicx}% Include figure files
\usepackage{dcolumn}% Align table columns on decimal point
\usepackage{bm}% bold math  
\usepackage[latin1]{inputenc}  
\usepackage[T1]{fontenc} 

\begin{document}

\preprint{AIP/123-QED}

\title{Room temperature operation of $n$-type Ge/SiGe terahertz quantum cascade lasers predicted by non-equilibrium Green's functions}
%\title{Ge/SiGe terahertz quantum cascade laser for room temperature operation investigated with a non-equilibrium Green's function model [Improving temperature robustness of terahertz quantum cascade laser using group IV materials]}

\author{T. Grange}
 \affiliation{nextnano GmbH, Lichtenbergstr. 8, Garching b. M\"unchen, 85748, Germany}
\author{D. Stark}
 \affiliation{Institute for Quantum Electronics, ETH Zurich, Zurich, 8093, Switzerland}
\author{G. Scalari}
 \affiliation{Institute for Quantum Electronics, ETH Zurich, Zurich, 8093, Switzerland}
\author{J. Faist}
 \affiliation{Institute for Quantum Electronics, ETH Zurich, Zurich, 8093, Switzerland}
\author{L. Persichetti}
 \affiliation{Dipartimento di Scienze, Universit\`a di Roma Tre, Roma, 00146, Italy}
\author{L. Di Gaspare}
 \affiliation{Dipartimento di Scienze, Universit\`a di Roma Tre, Roma, 00146, Italy}
\author{M. De Seta}
 \affiliation{Dipartimento di Scienze, Universit\`a di Roma Tre, Roma, 00146, Italy}
\author{M. Ortolani}
 \affiliation{Dipartimento di Fisica, Universit\`a di Roma ``La Sapienza'', 00185 Rome, Italy }
 \author{D.J. Paul}
 \affiliation{School of Engineering, University of Glasgow, Glasgow, G12 8LT, UK}
  \author{G. Capellini}
 \affiliation{IHP - Leibniz-Institut f\"ur innovative Mikroelektronik, Im Technologiepark 25, D-15236 Frankfurt (Oder), Germany}
  \affiliation{Dipartimento di Scienze, Universit\`a di Roma Tre, Roma, 00146, Italy}
 \author{S. Birner}
 \affiliation{nextnano GmbH, Lichtenbergstr. 8, Garching b. M\"unchen, 85748, Germany}
 \author{M. Virgilio}
 \affiliation{Dipartimento di Fisica ``E. Fermi'', Universit\`a di Pisa, Pisa, 56127, Italy}
 
%\email{Second.Author@institution.edu.}

\date{\today}% It is always \today, today,
             %  but any date may be explicitly specified

\begin{abstract}
$n$-type Ge/SiGe terahertz quantum cascade laser are investigated using non-equilibrium Green's functions calculations. We compare the temperature dependence of the terahertz gain properties with an equivalent GaAs/AlGaAs QCL design. In the Ge/SiGe case, the gain is found to be much more robust to temperature increase, enabling operation up to room temperature. The better temperature robustness with respect to III-V
is attributed to the much weaker interaction with optical phonons. The effect of lower interface quality is investigated and can be partly overcome by engineering smoother quantum confinement via multiple barrier heights.
\end{abstract}

%\pacs{Valid PACS appear here}% PACS, the Physics and Astronomy
                             % Classification Scheme.
%\keywords{Suggested keywords}%Use showkeys class option if keyword
                              %display desired
\maketitle

Terahertz (THz) quantum cascade lasers (QCLs) have been demonstrated with different III-V materials including GaAs/AlGaAs \cite{kohler2002terahertz}, InGaAs/AlInAs  \cite{ajili2005ga, deutsch2010terahertz}, InGaAs/GaAsSb \cite{deutsch2012high} and InAs/AlAsSb \cite{brandstetter2016inas}. 
In the past decade however, relatively small progress has been reported to increase the maximum operating temperature (presently 200~K) despite substantial efforts of design optimization \cite{kumar2009186, fathololoumi2012terahertz, franckie2018two}. The rationale for the quenching of THz laser emission above this temperature is due to the very effective electron--phonon (e--phonon) interaction, typical of III-V materials.
Indeed in polar lattices the longitudinal optical (LO) phonons induce a long-range polarization field which strongly couples to the charge carriers (Fr\"ohlich interaction). 
The THz transitions are typically designed to be well below the optical phonon energy (30--36 meV), so that at low temperature the upper laser state is protected against scattering by emission of LO-phonons. With increasing temperature however, the thermally activated electrons in the subband of the upper lasing state gain enough in-plane kinetic energy to access this scattering channel \cite{williams2007terahertz}.
This non-radiative relaxation of carriers reduces the population inversion and is responsible for quenching of the laser emission with increasing temperature as the gain drops below the cavity losses.
This fast non-radiative channel cannot easily be overcome by design engineering, as there is a necessary trade-off between selectivity and speed of the carrier injection in such closely spaced energy levels.

To overcome this limitation,  QCLs based on crystals having large optical phonon energy such as GaN or ZnO have recently been proposed\cite{wang2018broadening}. 
As an alternative strategy,
non-polar material systems are attractive because of their weaker e--phonon interaction. 
Indeed in these crystals the e--phonon coupling is controlled by the deformation potential which due to its short range is much less effective than the Fr\"ohlich interaction.
%As a consequence one can expect that optical-phonon scattering may not be the limiting factor in suppressing the gain at room temperature.  
SiGe alloys fulfill this requirement and also have the great advantage of being non-toxic materials fully compatible with silicon technology.
Among different configurations (electron or hole based, Si or Ge rich regimes) \cite{friedman2001sige,lynch2002intersubband,lever2008simulated,matmon2010si}, theoretical studies have indicated $n$-type Ge/SiGe heterostructures where charge transport is associated to L electrons, as the most promising architecture\cite{driscoll2007design,valavanis2008theory,valavanis2011material}.
Experimentally, sharp THz absorption peaks, related to intersubband transitions in $n$-type strain compensated Ge/SiGe quantum wells (QWs) have been demonstrated in the 20--50 meV region\cite{de2009conduction,busby2010near,de2012narrow} which interestingly covers the Reststrahlen band of III-V compounds. 
Moreover, long subband lifetimes have been evidenced with a weak dependence on lattice temperature \cite{ortolani2011long,virgilio2014combined,Ortolaniacsphotonics2016,VirgilioPRB2012}, and precise control of carrier tunnelling between coupled QWs has been demonstrated \cite{CianoPRASub}.
%Moreover, THz pump-probe experiments performed with Ge/SiGe QWs evidence very long subband lifetimes, with a weak dependence on lattice temperature \cite{ortolani2011long,virgilio2014combined,Ortolaniacsphotonics2016,VirgilioPRB2012}.
%Recently, precise control of carrier tunnelling of L electrons between adjacent coupled Ge/SiGe wells has been demonstrated \cite{CianoPRASub}.

THz gain in $n$-type Ge/SiGe QCL structures has been previously predicted using rate equation methods \cite{driscoll2007design, valavanis2008theory,lever2009importance} and a density matrix formalism \cite{dinh2012extended}. 
However, (i) dephasing effects were either not accounted (rate equation models), or described with phenomenological parameters (density matrix), while (ii) the effective electron temperature was entering the models as a free external input parameter.
Yet, dephasing in THz QCLs is a crucial issue since linewidths are comparable to the energy separation between the laser levels. 
In addition, a proper treatment of the transport-induced carrier heating effects is of paramount importance in Ge/SiGe systems, due to the low rate of energy transfer from the electronic to the phononic degrees of freedom, associated to the weak e--phonon interaction.
In this regard, more predictive calculations can be expected from the non-equilibrium Green's functions (NEGF) formalism since
(i) it does not require a phenomenological description of dephasing, as all the scattering processes are directly calculated from the material parameters;
(ii) there is no need of \textit{a priori} assumption for the in-plane electron distribution,
as carrier heating is accounted for in a self-consistent way.

In this work, to assess the potential of the SiGe alloy material system as a gain medium for intersubband cascade devices, we use the NEGF formalism to benchmark
a Ge/SiGe 4-quantum well QCL against a GaAs/AlGaAs counterpart \cite{amanti2009bound}.
To this aim, we preliminary validate our model comparing simulated
threshold current densities at different lattice temperatures with experimental data obtained with the GaAs/AlGaAs device, achieving good agreement.
Our main findings confirm that Ge/SiGe devices, although featuring at low temperature a reduced material gain with respect to III-V systems, are much less sensitive to an increase in temperature.
As a consequence we predict that, leveraging on efficient waveguides with optical losses not larger than 25~cm$^{-1}$, room temperature operation can be achieved in multilayer systems with interface roughness (IFR) lower than 2~\AA.

%The application of the NEGF formalism to understand vertical transport and gain properties in QCLs has been pioneered by Lee and Wacker\cite{lee2002nonequilibrium} and subsequently adopted also to investigate optical emission in the THz range\cite{nelander2008temperature,kubis2009theory,franckie2018two}.
The NEGF formalism has been shown to provide a powerful framework for investigating vertical transport and gain properties in QCLs \cite{lee2002nonequilibrium,nelander2008temperature,kubis2009theory,franckie2018two}.
Here, we perform NEGF calculations using the nextnano.QCL simulation package based on the model described in Refs.~\onlinecite{grange2014electron,grange2015contrasting} where scattering by acoustic and optical phonons, charged impurities, IFR, and alloy disorder have been accurately modeled by taking into account the full dependence of the scattered in-plane momentum.
In addition e--e scattering is included in a self-consistent one-particle elastic approximation, assuming that the calculated carrier density represents a fixed charge distribution for Coulomb scattering \cite{lin2018optimization}.
For the Ge/SiGe material system with [001] growth direction, we consider electron transport as due to carriers belonging to the fourfold degenerate L-valleys. 
Interaction with optical phonons in non-polar materials is controlled by the deformation potential only and in our model it is accounted using the same value of $3.5 \times 10^8$~eV/cm to describe both intra- and inter-valley events; furthermore in both the cases we use an effective dimensionless branch at 37~meV\cite{virgilio2014combined}.
%We stress that with the NEGF formalism it is not required to introduce an external parameter for the electron temperature, since the non-equilibrium distribution (for both growth and in-plane directions) is calculated in a self-consistent way from the transport and scattering processes.

\begin{figure}
\includegraphics[width=0.48\textwidth]{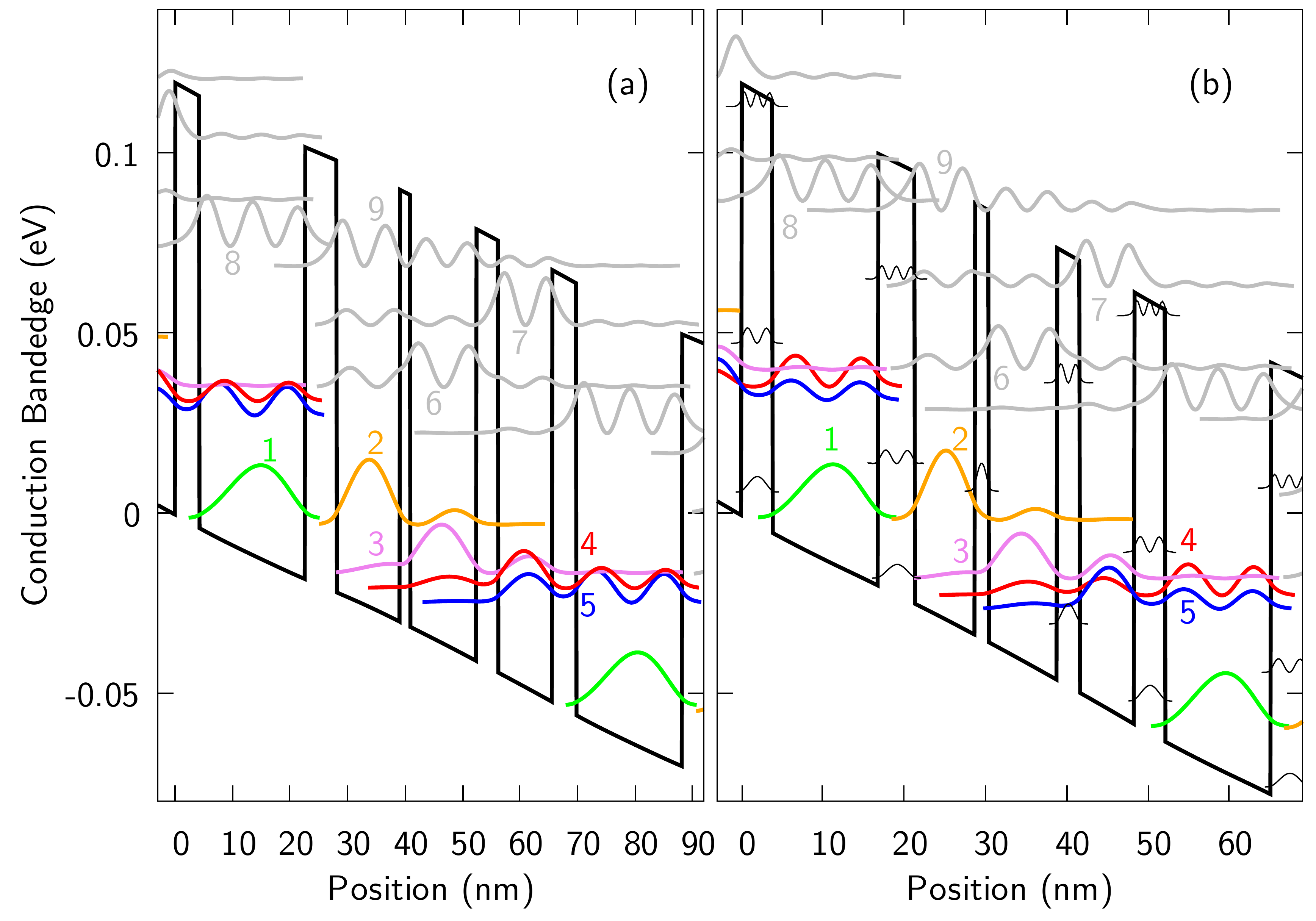}
\caption{\label{fig:WS} The conduction band profile and electronic states (squared modulus) for the four-well GaAs/Al$_{0.15}$Ga$_{0.85}$As (a) and Ge/Si$_{0.23}$Ge$_{0.77}$ (b) QCL design calculated for applied electric fields of 7.9 and 12.0 kV/cm, respectively. The electronic states shown are solutions of the Schr\"odinger equation on a single period limited by the thicker tunneling barrier
%separated by the larger SiGe barrier through which resonant tunneling occurs between states 1 and 2
(tight-binding basis).
Active levels are indicated in color and labelled from 1 to 5. Higher energy states accounted in the NEGF simulations are shown in light grey lines.
In panel (b) $\Delta_2$ states confined in the barriers are also shown (dark grey lines).}
\end{figure}

To compare the performance of SiGe based THz QCL devices against their III-V counterpart, we choose as a reference the four-well bound-to-continuum design introduced in Ref.~\onlinecite{amanti2009bound} using the GaAs/AlGAs material system.
Our choice is motivated by the scalability of this design in term of emission frequency and by its robustness against deviations of the layer thicknesses or concentrations from the nominal values, which made it very suited for heterogeneous cascade device \cite{rosch2015octave}.
As shown in Fig.~\ref{fig:WS}(a) carrier injection is based on resonant tunneling from level 1 to level 2; the lasing transition occurs from this latter state to level 3, while levels 4 and 5 act as continuum to extract the carrier from the lower laser level. 
Finally relaxation in the injector state involves resonant emission of optical phonons (5 to 1 of next period). 
As shown in Fig.~\ref{fig:WS}(b) a very similar electronic spectrum can be engineered in a QCL architecture based on a Ge/Si$_{0.23}$Ge$_{0.77}$(001) heterostructure, which features a comparable band-offset, properly adjusting layer thicknesses to account for the heavier confinement L mass (0.12 $m_0$) \cite{driscoll2007design}.
Minimizing the elastic energy associated to the tensile (compressive) SiGe barriers (Ge wells) we find that this system matches the strain-balance conditions when a relaxed Ge/Si$_{0.055}$Ge$_{0.945}$ virtual substrate is adopted.
Note that this strain field splits the sixfold $\Delta$ degeneracy, lowering (rising) the energy of the two valleys located along the growth direction in the barrier (well) region.
It follows that $\Delta_2$ states well confined in the Si$_{0.23}$Ge$_{0.77}$ layers are present in the  same energy region as the relevant L states (see Fig.~\ref{fig:WS} (b)).
We expect that these states play a minor role in the transport process, due to their strong confinement.
Moreover intersubband transitions between $\Delta_2$ levels occur at frequencies much larger than the one involving level 2 and 3.
For these reasons, we expect that those states  have a limited impact on the gain spectra 
and therefore we neglect their presence in the NEGF simulations.

%To have an intuitive idea of the relevant differences between the GaAs/AlGaAs and Ge/SiGe material system as active medium in QCL devices, in Table~\ref{table} we compare the literature values of some material parameters. 
Three relevant  material parameters differ notably between the GaAs/AlGaAs and  Ge/SiGe material systems (see the supplementary  material).
First, the effective mass along the growth direction is higher in Ge (0.12) than in GaAs (0.07),
%reducing the tunneling rates and being
which is detrimental for optical amplification since
dipole matrix elements scale as $m^{-1/2}$.
%with a fixed carrier distribution in the laser levels, the gain scales as $m^{-1}$. 
Second, as already noticed the interaction with optical phonons is much weaker in Ge as optical lattice excitations do not induce long-range polarization fields\cite{virgilio2014combined}.
Third, the dielectric constant is higher in Ge (16.2) than in GaAs (12.9). 
This has a favorable consequence, as the elastic scattering rate due to Coulomb interaction (e-impurity and e--e interactions), which has been identified as the principle source of dephasing in GaAs/AlGaAs THz QCLs\cite{grange2015contrasting}, scales as the inverse square of the dielectric constant.

\begin{figure}
\includegraphics[width=0.4\textwidth]{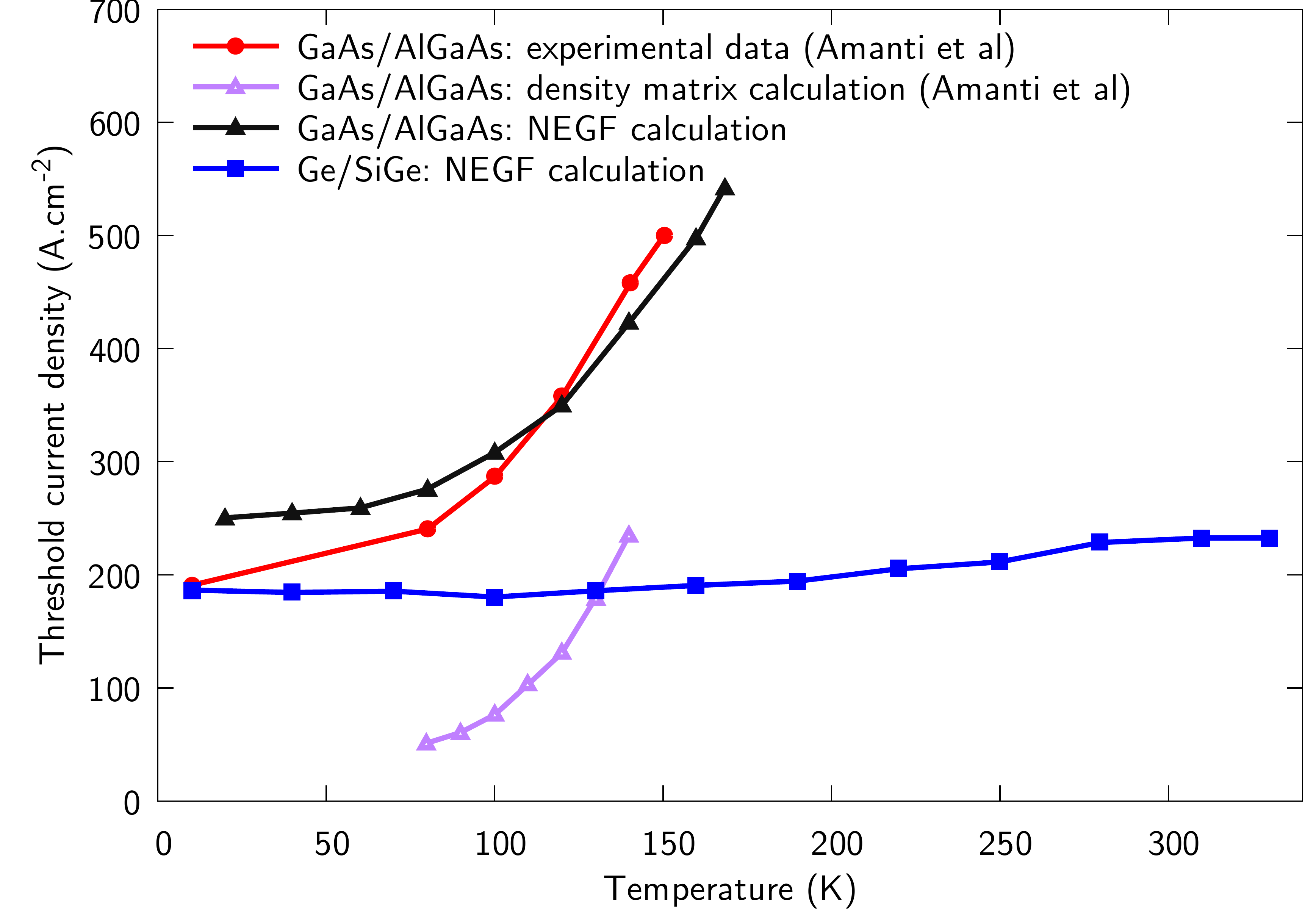}
\caption{\label{fig:threshold} The threshold current density as a function of lattice temperature for the four-well Ge/SiGe QCL (blue squares) and the GaAs/AlGaAs design. For the III-V device experimental measurements from Ref.~\onlinecite{amanti2009bound} (red circles) are compared with theoretical data based on a density matrix model reported therein (purple open triangles) and with the results of our NEGF model (black filled triangles). Cavity losses of 25~cm$^{-1}$ and IFR rms of 1~\AA are assumed.}
%NEGF simulations have been obtained assuming cavity losses of 25~cm$^{-1}$ and IFR rms of 1~\AA.}
\end{figure}

To validate the predictivity of our NEGF approach we compare in Fig.~\ref{fig:threshold} the calculated threshold current as a function of temperature for the GaAs/AlGaAs device with the experimental data reported in Ref.~\onlinecite{amanti2009bound}, assuming cavity losses of 25 cm$^{-1}$. A much better agreement is obtained compared to numerical results based on the density matrix model developed by the same authors.  
It is worth noticing that no fitting parameters are used, the IFR scattering playing only a minor role since
very similar values for the threshold current are obtained completely neglecting this scattering channel (not shown).
%As discussed in the following, this is not the case for the Ge/SiGe device.
For the Ge/SiGe device, assuming the same IFR root-mean square (rms) value of 1~\AA{},
%, beside obtaining lower current threshold densities
we observe an almost negligible variation of threshold current densities as a function of the temperature, with an operating range extending up to 330~K.
%This strongly support the potential of the Ge/SiGe material system for room temperature operation.   

\begin{figure}
\includegraphics[width=0.4\textwidth]{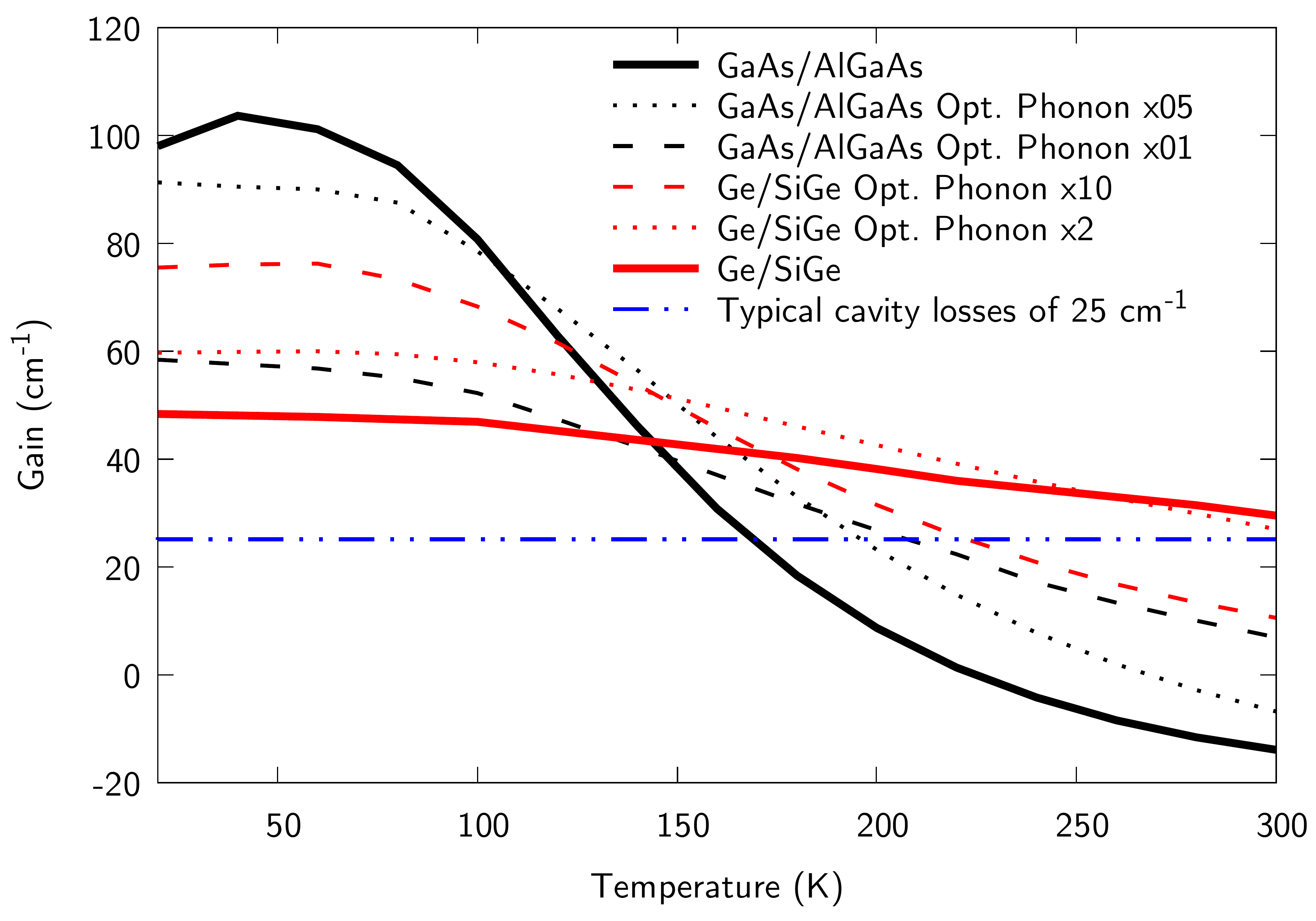}
\caption{\label{fig:Tdep} (Color online) The maximum gain as a function of temperature calculated for the GaAs/AlGaAs design of Amanti et al. (black solid line) and the proposed Ge/SiGe design (red solid line). 
The black dot and dash lines have been obtained for the GaAs/AlGaAs QCL scaling the square of the optical-phonon coupling constant by 0.5 and 0.1, respectively. 
The red dot and dash lines refer to the Ge/SiGe QCL with a scaling for the square of the optical-phonon coupling of 2 and 10, respectively. 
The blue dash-dot-dot horizontal line at 25 cm$^{-1}$ represents a typical value for cavity losses.}
\end{figure}

%To better appreciate this point, 
In Fig.~\ref{fig:Tdep} we compare the value of the peak material gain as a function of temperature in the GaAs/AlGaAs and Ge/SiGe QCLs. 
At low temperatures, the peak gain is higher in the III-V device. 
In this case however, the gain rapidly drops as the temperature increases.
Assuming cavity losses of 25~cm$^{-1}$, this drop leads to a maximum operation temperature of 168~K, in good agreement with the experimental value of 150~K reported in Ref.~\onlinecite{amanti2009bound}.
Conversely, the maximum gain for the Ge/SiGe QCL displays a different behavior, being weaker at low temperature but much more robust against the temperature increase. 
In line with the above results, the predicted gain at 300~K, although reduced to 30~cm$^{-1}$, remains larger than the cavity losses, thus maintaining the laser emission at room temperature.

This remarkable difference in the temperature dependence of the two QCL devices can be attributed to the much weaker e--phonon interaction of non-polar lattices, as demonstrated below by artificially tuning the e-optical-phonon coupling constant in the two systems.
To this aim in Fig.~\ref{fig:Tdep} we report the peak gain obtained scaling the square of the e-optical-phonon coupling so to suppress (enhance) the scattering rate in the GaAs/AlGaAs (Ge/SiGe) QCL. 
At low temperature a weaker e--phonon interaction in the III-V based device diminishes the gain but increases it in the high temperature region while the opposite happens upon increasing the interaction in the Ge/SiGe system.
This behavior can be understood by the double role played by optical phonons. 
In fact, for what concerns the lasing transition, increasing the optical-phonon scattering rate is detrimental for the population inversion, especially at high temperature since the scattering from the upper to the lower laser subband can be efficiently activated by the thermal energy.
On the other hand, one has to consider that scattering by optical phonons also controls the relaxation from the lower laser level to the injector state of the next period. 
It follows that when the scattering is reduced, the relaxation rate towards the injector level, which in the design considered is based on optical phonon emission becomes slower and this fact, which is the dominant effect at low temperature, negatively impacts the gain. 

\begin{figure}
\includegraphics[width=0.49\textwidth]{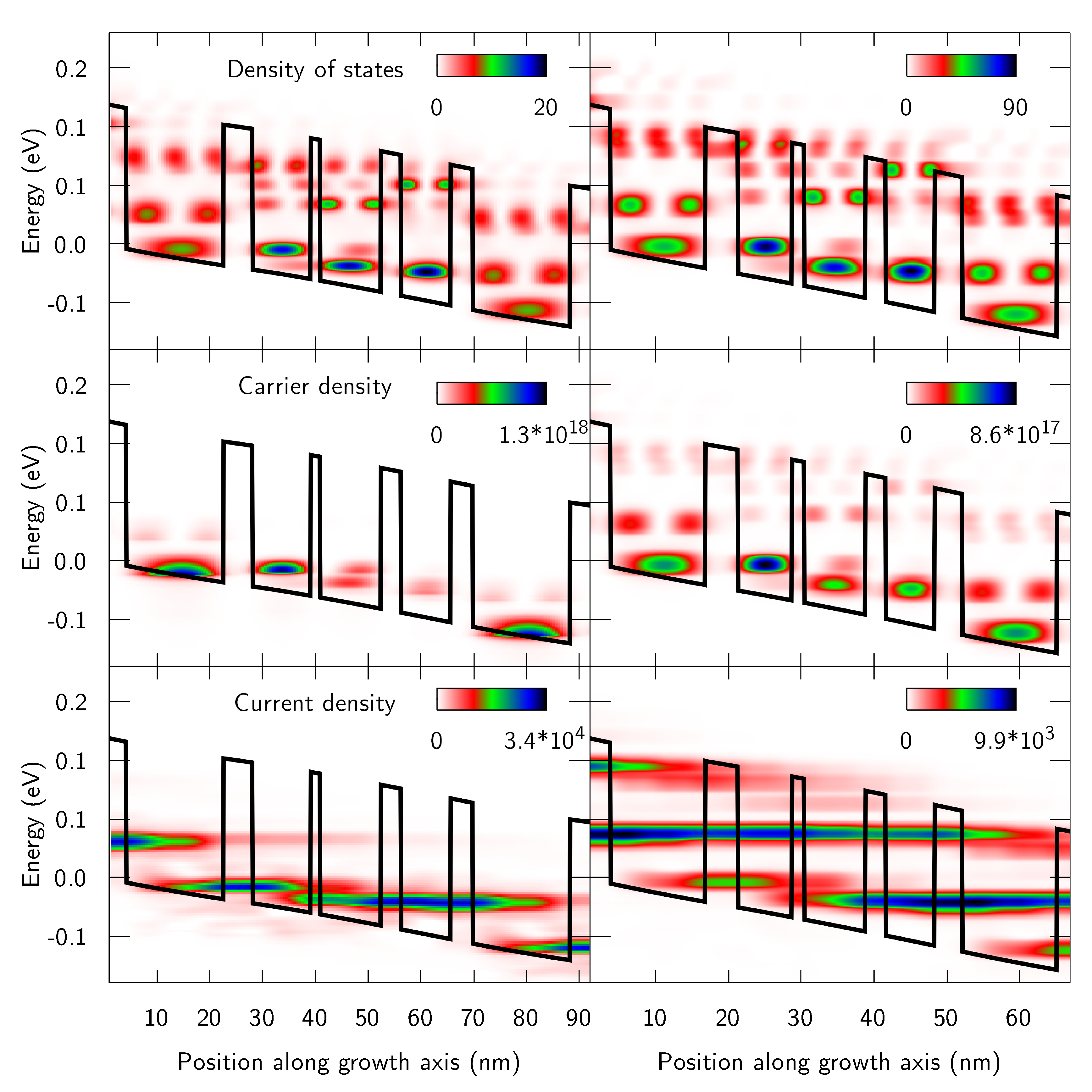}
\caption{\label{fig:2D} Position and energy-resolved 1D density of states (top) in (nm$^{-1}$eV$^{-1}$),  carrier density (middle) in (cm$^{-3}$eV$^{-1}$) and current density (bottom) in (Acm$^{-2}$eV$^{-1}$) for the GaAs/AlGaAs design (left) and for the Ge/SiGe design (right).}
\end{figure}

\begin{figure}
\includegraphics[width=0.49\textwidth]{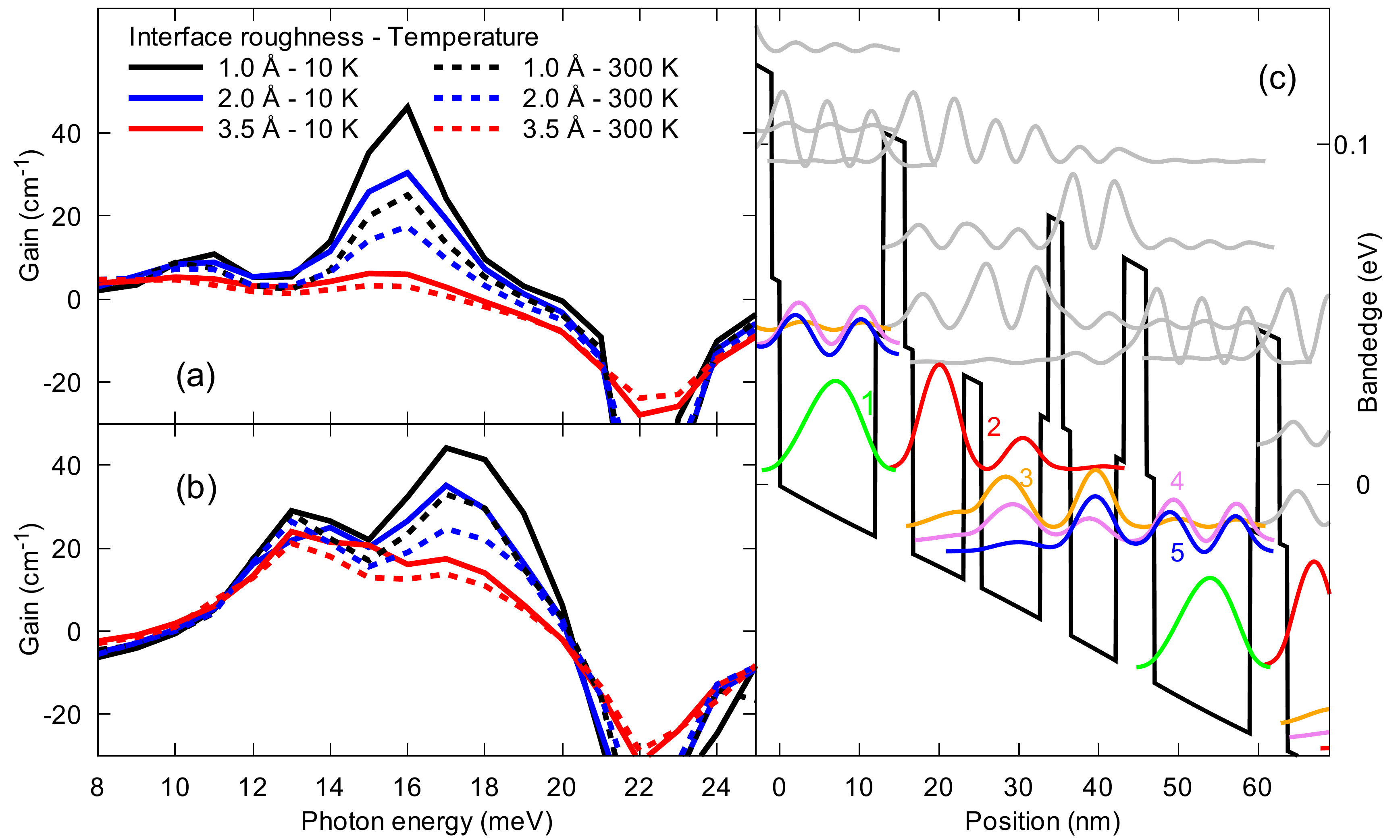}
\caption{\label{fig:roughness} Material gain spectra for the Ge/SiGe QCL device with single barrier height (a) and multiple barrier heights (b) at lattice temperature of 10~K (solid lines) and 300~K (dotted lines) for various interface roughness rms deviation at constant correlation length (70~\AA). The modified design with multiple barrier heights is displayed in (c). }
\end{figure}

In Fig.~\ref{fig:2D} we compare the position and energy-resolved density of states (top), carrier density (middle) and current density (bottom) calculated for the GaAs/AlGaAs and the Ge/SiGe QCL at the electrical bias corresponding to the peak gain for T=10 K.
From the top panels it is apparent that the electronic spectra are very similar.
Nevertheless, carrier population of the higher energy levels is negligible in the GaAs/AlGaAs case only (middle panels).
This fact has a relevant impact on the current distribution since the higher-energy occupied states contribute significantly to the charge transport as shown in the bottom panels.
This effect is related to the much less efficient carrier thermalization in the Ge/SiGe device, caused by the weaker interaction with the lattice degrees of freedom. 
As a consequence, the excess electron effective temperature, evaluated from the in-plane population distribution, is found to be much higher (240 K) than in the Ga/AlGaAs device (110 K). 
Another effect related to the large effective temperature is the presence of parasitic absorption peaks associated to thermally activated transitions, involving carriers belonging to the higher energy subbands.
In the gain spectrum shown in Fig.~\ref{fig:roughness}(a), a dip is calculated at a photon energy of 22~meV owing to the absorption from level 6 to level 7.
This absorption feature is not observed in the GaAs/AlGaAs case because of the much smaller population of level~6. In contrast, in the Ge/SiGe case, this parasitic absorption has to be accounted in the design to prevent overlapping with the gain peak. To this aim, we have adopted QWs which are narrow enough to push this absorption line to an energy (22~meV) well above the one of the lasing transition (16~meV).
This issue, which has not been considered in previous studies of Ge/SiGe QCLs \cite{lever2009importance}, should be carefully taken into account when optimizing a Ge/SiGe QCL design.

%, with an absorption line occurring at 22~meV in Fig~\ref{fig:roughness}(a). 

%Due to the weaker e--phonon coupling, the IFR scattering plays a more relevant role in QCLs based on the Ge/SiGe material system.
%Therefore a precise description of this physical effect is crucial for predictive modelling.
The role played by IFR in the Ge/SiGe system is investigated below, as the interface quality in Ge/SiGe multilayers grown by CVD could be lower than the typical one achievable for GaAs/AlGaAs structures grown by MBE \cite{califano2007interwell, busby2010near}.
Indeed, results presented so far have been obtained assuming the same IFR parameters for GaAs/AlGaAs and Ge/SiGe QCLs, namely a rms deviation in the growth direction of 1~\AA{} and an in-plane correlation length of 7~nm. 
To shed light on the role of IFR, in Fig.~\ref{fig:roughness}(a) we have calculated the gain spectra with different values of the rms deviation from the ideal interface for the Ge/SiGe QCL introduced above (left panel).
As the IFR increases, the peak gain decreases at both low and high temperatures. 
Again, setting cavity losses at 25 cm$^{-1}$, we find that an IFR rms deviation of 2~\r{A} enables operation only at low temperatures, while a larger value of 3.5~\r{A} prevents any lasing action.

To mitigate the effect of IFR, more advanced designs can be envisaged \cite{bai2010highly,semtsiv2012low}.
%Since this optimization is beyond the scope of the present work, we limit here to compare the gain robustness against IFR of the Ge/SiGe QCLs device with 
We propose in Fig.~\ref{fig:roughness}(c) a modified design in which smoother interfaces are engineered by adding layers with lower Si concentration (Si$_{0.11}$Ge$_{0.89}$).
%In order to further simulate our structure and understand possible limitations coming from IFR, we compare the gain robustness against IRS of the Ge/SiGe QCLs device with a modified design in which we have engineered smoother confinement. To this purpose,
A 1~nm thick layer of Si$_{0.11}$Ge$_{0.89}$ is introduced at each interface between Ge and Si$_{0.23}$Ge$_{0.77}$.  The barrier separating the two laser levels is replaced by a thicker Si$_{0.11}$Ge$_{0.89}$ barrier. 
The 5 active levels of the QCLs having energies lower than the intermediate barrier height, their probability density at the interface with the Si$_{0.23}$Ge$_{0.77}$ barrier material is reduced, owing to the evanescent decay of their wavefunctions in the intermediate barrier layer.
In this way these active levels mainly interact with the IFR contact potential associated to a reduced barrier height of 60~meV instead of 120~meV and then lower IFR scattering rates are expected.
Optical gain for this improved design is shown in Fig.~\ref{fig:roughness}(b). The spectra are found to be more robust against IFR, confirming that such or similar design strategies can be pursued to overcome the detrimental effects associated to large IFR.

In summary, using NEGF simulations, we have assessed the potential of Ge/SiGe material system to achieve THz emission at room temperature in QCL devices through a detailed comparison with an equivalent GaAs/AlGaAs design.
%after having benchmarked the predictivity of our model with  experimental data reported in the literature for a III-V device,
The Ge/SiGe QCL is found to be significantly more robust as the temperature increases,  %, provided that the IFR is lower than 2~\AA.
which is clearly attributed to the weaker electron--phonon interaction. 
% which is also responsible for the observed higher effective electron temperatures. 
Finally, we have shown that detrimental effects related to possible high IFR can be attenuated by engineering a smoother confinement profile adopting a three layer barrier. % featuring a triangular shape.
We believe that the present results will motivate new experimental efforts aimed at demonstrating room-temperature operation in group IV QCL THz devices.

\begin{acknowledgments}
This project has been funded by the European Union's Horizon 2020 research and innovation programme under grant agreement No 766719 (FLASH).
\end{acknowledgments}

\appendix

\end{document}